\documentclass[superscriptaddress,amsmath,amssymb,aps,prb,onecolumn,reprint,floatfix]{revtex4}
\pdfoutput=1
\usepackage{amssymb,amsmath,bbm,braket,indentfirst,bm}
\usepackage{dcolumn}
\usepackage{color}
\usepackage{bigstrut}
\usepackage{graphicx}
\usepackage{framed}
\usepackage{setspace}
\usepackage{lineno}
\usepackage[colorlinks,linkcolor={blue},citecolor={blue},urlcolor={blue}]{hyperref}
\begin{document}

\title{Cellular reprogramming dynamics follow a simple one-dimensional reaction coordinate}

\author{Sai~Teja~Pusuluri}
\affiliation{Department of Physics and Astronomy and Nanoscale and Quantum Phenomena Institute, Ohio University, Athens, OH, 45701, USA}
\affiliation{These authors contributed equally to this work.}

\author{Alex~H.~Lang}
\affiliation{Physics Department, Boston University, Boston, Massachusetts 02215, USA}
\affiliation{Center for Regenerative Medicine, Boston University, Boston, MA, 02215}
\affiliation{These authors contributed equally to this work.}

\author{Pankaj~Mehta}
\affiliation{Physics Department, Boston University, Boston, Massachusetts 02215, USA}
\affiliation{Center for Regenerative Medicine, Boston University, Boston, MA, 02215}
\affiliation{Co-corresponding author: pankajm@bu.edu}

\author{Horacio~E.~Castillo}
\affiliation{Department of Physics and Astronomy and Nanoscale and Quantum Phenomena Institute, Ohio University, Athens, OH, 45701, USA}
\affiliation{Co-corresponding author: castillh@ohio.edu}
\date{\today}
\begin{abstract} 
Cellular reprogramming, the conversion of one cell type to another, has fundamentally transformed our conception of cell types. Cellular reprogramming induces global changes in gene expression involving hundreds of transcription factors and thousands of genes and understanding how cells globally alter their gene expression profile during reprogramming is an open problem.  Here we reanalyze time-series data on cellular reprogramming from differentiated cell types to induced pluripotent stem cells (iPSCs) and show that gene expression dynamics during reprogramming follow a simple one-dimensional reaction coordinate. This reaction coordinate is independent of both the time it takes to reach the iPSC  state as well as the details of experimental protocol used.  Using Monte-Carlo simulations, we show that such a  reaction coordinate emerges naturally from epigenetic landscape models of cell identity where cellular reprogramming is viewed as  a ``barrier-crossing'' between the starting and ending cell fates. The model also provides gene-level insight into reprogramming dynamics and resolves a debate in the stem cell field about the different phases of reprogramming dynamics. Overall, our analysis and model suggest that gene expression dynamics during reprogramming follow a canonical trajectory consistent with the idea of an ``optimal path'' in gene expression space for reprogramming.
\end{abstract}

 \maketitle


Biology is in the midst of the revolution spearheaded by the pioneering work of Takahashi and Yamanaka\cite{Takahashi2006Induction} on cellular reprogramming showing that it is possible to reprogram mouse embryonic fibroblasts (MEFs) to cells resembling embryonic stem cells (ESCs), commonly called induced pluripotent stem cells (iPSCs), by manipulating the expression of just four transcription factors (TFs). The idea of manipulating small sets of TFs to alter cell fates has proven extremely versatile and  it is now possible to create iPSCs  from a variety of cell types\cite{Gonzalez2011Methods}, as well as perform direct conversions between two differentiated cell types such as MEFs and neurons \cite{Vierbuchen2010Direct}. Most reprogramming experiments  have a similar design \cite{Takahashi2007Induction-2} (Fig 1A). The starting cell type (e.g.~MEF) is engineered with a construct containing the desired reprogramming genes. These genes are induced at the start of the experiment. After several days, the cell culturing conditions are switched to a medium favorable to the desired cell type (e.g.~ stem cell media). At a later time, typically a  few weeks, the exogenous genes are turned off. If all goes well, a small percentage ($\approx 0.01 - 1\%$) of cells successfully  reprogram to the desired cell type. 

Significant progress has been made towards understanding the mechanisms underlying cellular reprogramming\cite{Yamanaka2012Induced,Xu2015Direct} (which from now on we will use to include both reprogramming to iPSC as well as direct conversion), yet many questions remain. Cellular reprogramming requires global changes in gene expression involving hundreds of transcription factors and thousands of genes, but how cells dynamically alter their gene expression profile during reprogramming is still not well understood. Reprogramming rates seem to depend on the exact protocol used and can be changed by several orders of magnitude through careful genetic manipulations\cite{Hanna2009Direct,Rais2013Deterministic}. Experiments have also measured whole genome time courses during reprogramming but the high-dimensional nature of the measured trajectories make them difficult to interpret \cite{Polo2012A-Molecular}. Other experiments have examined gene-level events during reprogramming. Buganim et al \cite{Buganim2012Single-Cell} analyzed reprogramming dynamics at the single-cell level and concluded that reprogramming initially is probabilistic but ends with a hierarchichal (i.e.~ordered), deterministic stage. In contrast, Polo et al  \cite{Polo2012A-Molecular} analyzed reprogramming dynamics with both population level and single-cell level measurements and concluded that reprogramming follows an early deterministic phase with many gene changes, followed by an intermediate phase with fewer changes, and ending with a deterministic phase with many gene changes. Recently, Chung et al \cite{Chung2014Single} measured single cell reprogramming dynamics and proposed that the intermediate phase of reprogramming is a ``loosely ordered probabilistic phase'' in which the timing between events is probabilistic, but the order of events is relatively deterministic. This  highlights the need for a better understanding gene expression dynamics during reprogramming. 

Reprogramming involves global changes in gene expression and hence is intrinsically high dimensional. For this reason, it is common to use dimensional reduction techniques such as Principal Component Analysis (PCA) to project the dynamics onto a low-dimensional sub-space. However, dimensional reductions techniques such as PCA  have several key limitations.  The principal component vectors have no clear biological interpretation, making it difficult to extract biological meaning from the resulting low-dimensional dynamics. PCA also depends  on the type and quality of the data included in the dataset, making it cumbersome to compare dynamical data across experiments and systems.

To overcome these challenges, we introduce a new technique for visualizing high-dimensional reprogramming dynamics inspired by ``epigenetic landscape'' models for cellular identity. In Waddinton's original landscape idea \cite{Waddington1957The-Strategy}, cell types correspond to basins of attraction in an abstract cell identity landscape. This idea has been refined by a variety of researchers, and has yielded a number of insights into the genetic basis of cellular identity \cite{Kauffman1993The-Origins,Enver2009Stem,Huang2009Reprogramming,Wang2010The-Potential,Zhou2011Understanding,Zhou2011Predicting,Wang2011Quantifying-Waddington,Huang2012The-molecular,Li2013Quantifying,Banerji2013Cellular,Heinaniemi2013Gene-pair,Xu2014Exploring,Li2014Landscape,Zhang2014Stem}. Two of us recently proposed a landscape model \cite{Lang2014Epigenetic} that takes global gene expression profiles (microarrays or RNA-Seq) and uses techniques inspired by spin physics and the Hopfield model to explicitly construct a cell identity landscape. This model  provided a natural explanation for the existence of partially-reprogrammed cell types and  can identify TFs that have been used to successfully reprogram to multiple cell types. In this paper, we extend our previous work to analyze reprogramming dynamics. Using a new linear-algebra based analysis method inspired by our landscape model, we show that the experimentally observed gene expression dynamics during reprogramming follow a simple, one-dimensional reaction coordinate.  This reaction coordinate emerges naturally in numerical simulations of our landscape model, suggesting that reprogramming can be understood as  a  ``barrier crossing'' between landscape minima.

\section{Results}

\subsection{Mathematical Model and Data Analysis Method}
Here,  we briefly summarize the relevant features of the landscape model (see Materials and Methods and Lang et al. \cite{Lang2014Epigenetic} for details). Cell types are stable basins of attraction (minima of the landscapes) and reprogramming  between basins proceeds through stochastic fluctuations resulting from gene expression noise (Fig 1B). The landscape is constructed directly from the genome wide expression profiles of natural cell types  using a curated dataset of microarrays for $p=63$ cell types and approximately $N \sim 1400$ TFs (see Materials and Methods). This data is summarized in a cell type matrix, $\xi^\mu_i$, whose entries contain the expression level of  TF $i$ in cell type $\mu$ (e.g. MEF, ESC). This construction can easily be extended to include genes beyond TFs.

The global gene expression level of TFs in an arbitrary gene expression state can be summarized using a $N$-dimensional expression state vector $S_i$ whose entries encode the expression level of TF $i$ with $i=1\ldots N$. Expression levels are treated as continuous variables when analyzing experimental data and as binary variables which can be either on or off ($S_i=\pm1)$ when performing numerical simulations (see Materials and Methods). To analyze experimental data, it is useful to define a ``distance'' measure between an arbitrary expression vector $S_i$ and the expression vector, $\xi^\mu_i$, for cell type $\mu$. One natural distance measure in gene expression space is the overlap or dot product, $m^\mu =1/N \sum_i S_i \xi^\mu_i$, which measures the correlations between $S_i$ and $\xi^\mu_i$. The overlap between cell type $\mu$ and state $S_i$ is $1$, $-1$, or $0$ for a perfectly correlated, anti-correlated, or uncorrelated states, respectively. In practice, the dot product is a poor measure of distance because cell types are highly correlated with each other. For example, blood cell types share a common core set of gene expression and thus B cells and T cells have a 87\% overlap in their gene expression profiles.

For this reason, it is useful to introduce an alternative measure of distance  we call projections, with $a^\mu$ denoting the projection of $S_i$ on the expression profile of cell type $\mu$. The projection has a simple geometric interpretation depicted in Fig 1C and is calculated by first projecting (ie casting a shadow) of $S$ onto the hyperplane defined by the $p$ cell types in the matrix $\xi$ (represented as the gray plane) and then measuring the distance to the cell type $\mu$ within this cellular subspace. The benefit of this construction is that it naturally accounts for the correlations between cell types: the projection of a B cell with itself is one, while a B cell's projection on T cells is zero, and vice versa. This is in stark contrast with correlation based measure of distance in gene expression space. 

Projections arise naturally when constructing landscape models for cellular identity. In  Lang  et. al \cite{Lang2014Epigenetic}, it was shown that it is possible to define a Lyapunov function (commonly called an energy), $H$, that characterizes the landscape. In terms of the projections $a^\mu$ and overlaps $m^\mu$, the energy or Lyapunov function takes the form (see  Materials and Methods and \cite{Lang2014Epigenetic}):
\begin{eqnarray}
H &=& H_{\text{basin}} + H_{\text{culture}}\nonumber\\
 &=& -\frac{N}{2}\sum_{\mu=1}^p m^\mu a^\mu  - \sum_{\mu=1}^p b^\mu a^\mu.
\label{Hamiltonian}
\end{eqnarray}
We emphasize that this Lyapunov function represents an abstract ``cellular identity energy surface'' characterizing the stability of cell states and cannot be directly related to metabolism or ATP consumption. In this expression, the first term $H_{\text{basin}}$ arises from the ``effective'' interaction between genes and ensures that all cell types $\mu=1 \ldots p$ are attractors of the dynamics that have large basins of attraction. This can be seen by noting that in a given cell type (say $\mu=1$),  $S_i= \xi_i^1$ and $a^1=m^1=1$, while the projection on all other cell types  is zero, $a^\nu=0$ ($\nu=2 \ldots p$). Plugging these results into Eq. \ref{Hamiltonian} shows that each cell type is a global minimum with energy $H_{min}=-\frac{N}{2}$. The second term $H_{\text{culture}}$ represents the stabilizing effect of the culturing conditions on a particular cell type. For example, during reprogramming when cells are initially grown in MEF culture, only $b^{MEF}\neq 0$, while later in reprogramming, after the media has been changed to ESC culture, only $b^{ESC}\neq 0$. Finally, to incorporate the fact that some transcription factors are overexpressed in the experiments (see Materials and Methods) the dynamics of the variables $S_i$ corresponding to over expressed TF are locked in the ``on'' state. 

\subsection{MEF Reprogramming Dynamics}

We begin by reanalyzing the experimentally available time series data on reprogramming in mice. Fig 1D shows the first two principal components (PC) for 10 different reprogramming trajectories from MEF to iPSC from multiple labs. In the analysis, we have included partially reprogrammed cells (PRC), which are novel cell states only found during incomplete reprogramming experiments. The plot shows dynamics projected onto the first two PCs, but in reality this system is high-dimensional and it takes 21 PCs to explain 80\% of the variation in the data (see SI Fig 1 for details). The PCA plot illustrates several important findings. First, reprogramming trajectories seem to group into two distinct clusters, and within each cluster, the starting points (day 0) and ending points (final iPSC) are near each other. Therefore, even for different experimental protocols, reprogramming seems to follow only a few paths. Second, these paths are distinct from partially reprogrammed cells (PRC). While several reprogramming data points seem to be near PRCs, this is an artifact of keeping only two PCs in our visualization. In fact, the PRCs only have a Spearman correlation of 90\% with the closest reprogramming data point and approximately 80\% correlation with the two closest trajectories. Third, the final state of failed trajectories (trajectories that did not successfully reprogram to iPSCs) is closer to their starting point rather than to iPSCs, suggesting that failed trajectories do not leave the basin of attraction of the initial cell type. While PCA allows easy visualization of the data,  the Principal Components have no clear biological meaning making it difficult to interpret the lower dimensional PCA dynamics.

In Figures 1E-1G, we have replotted the same time-series data as in the PCA plots using projections on the starting and ending cell type. As in the PCA plot, the various symbols represent the actual data, while the lines connecting data show the time order of experimental points.  In these plots,  the starting (ending) states for each trajectory are defined as the initial (final) time point for the corresponding experiment. When calculating projections, the start (end) states replace MEF (ESC) in our cell type matrix $\xi$. This allows us to plot each experiment against its own start and end points. This additional step is necessary because different experiment define MEFs and iPSs differently. 

The result of this analysis is shown in Fig 1E. In contrast to the PCA plot, which contained two clusters (Fig 1D), the reprogramming trajectories in the projected basis all follow a similar path. This suggest cells follow a simple one-dimensional reaction coordinate during reprogramming: a straight line joining the starting cell type with the ending cell type in projection space. This data collapse is more remarkable when considering the extreme heterogeneity in reprogramming rates across the plotted experiments. The Polo et al experiment\cite{Polo2012A-Molecular} represents a typical time course with reprogramming taking approximately two weeks, while Rais et al \cite{Rais2013Deterministic} is the fastest trajectory (8 days) and ST (Samavarchi-Tehrani et al)\cite{Samavarchi-Tehrani2010Functional} is the slowest trajectory in our dataset (30 days). 

 In order to better understand how trajectories with such different reprogramming rates can still follow the same coordinate, it is useful to extend the analysis to account for how reprogramming trajectories project on other cell types besides the starting and ending cell types. To do so, we introduce a new quantity,
\begin{equation}
a_{\perp} = \sqrt{\sum_{1 \le \nu \le p \atop \nu\neq \left(\text{start},\text{end}\right) } \left( a^\nu \right)^2},
\end{equation}
that measures the magnitude of the projections perpendicular to the plane spanned by the starting and ending cell type. This is shown in Fig 1F. Notice that faster trajectories have a smaller perpendicular projection on the remaining cell types than slower trajectories. Furthermore, the difference in speed between experiments arises largely from the fact that slower trajectories also appear to get stuck at particular points along the reaction coordinate for as long as two weeks. 

To compare these experimental trajectories to our mathematical model, it is useful to visualize this data in yet another way. In Fig 1G, we have replotted the same data taking the z-axis as the energy per TF, which can be calculated directly from gene expression profiles using our landscape construction ($H_{\text{basin}}/N$). In making these plots we have ignored the contributions of the culture terms in Eq. \ref{Hamiltonian} to the energy in our model (see SI Figures and Material Methods).  Notice that the faster trajectories follow a lower energy path while the slowest trajectory (ST) follows a high energy path and appears to spend time stuck in two different barriers between days 8 and 21. These observation suggest that the experimentally observed reprogramming dynamics are consistent with the idea of a  ``barrier crossing'' between the starting and ending cell types in a rough landscape (see Fig 1B). 

Further evidence for this  barrier-crossing picture comes from numerical simulation using our landscape model (see Material Methods). The insets in Fig. 1E-1G show failed and successful reprogramming trajectories from Monte-Carlo simulations. There is a striking similarity between the model trajectories and experiment. Like in the experiments, successful reprogramming trajectories in our model  follow a simple one-dimensional reaction coordinate in the projection space and reprogramming requires crossing a significant energy barrier. Supplementary Figures 2-4 contain more examples of successful and unsuccessful simulation trajectories.

Finally, we note that the reaction coordinate can also be visualized using more traditional measures of distances  such as the overlap (dot product) of the gene expression profile with the starting and ending states (see SI Fig 5A). However, when using overlaps, each experiment has its own starting and ending point, making it hard to compare across experiments. Furthermore, overlaps are unable to discern the ``barrier crossing'' picture that emerges naturally from using projections (see SI Fig 5B).

\subsection{B Cell Reprogramming Dynamics}

The previous section considered reprogramming from MEF to iPSC. Here, we extend this analysis to consider two additional reprogramming experiments from B cells to IPSs\cite{Di-Stefano2014C/EBPalpha,Di-Stefano2014Time-resolved}. In the first experiment, the standard Yamanaka reprogramming protocol (OSKM)\cite{Takahashi2006Induction} was used to reprogram B cell to iPSC. Unlike in MEFs, in B cells the OSKM protocol resulted in extremely low reprogramming yields. To increase the reprogramming yield, the protocol was then modified so that OSKM expression was preceded by pulsed expression of CEBP$\alpha$  (abbreviated C+OSKM). This modified protocol significantly increased the reprogramming yield. Figure 2A shows that for both experiments, reprogramming trajectories once again follow a simple reaction coordinate in projection space. Figure 2B extends these plots to the energy vs  reaction coordinate plane. Notice, that in both experiments, the energy of the trajectories first increase and then decrease. The higher yield trajectory (C+OSKM) makes steady progress over the energy barrier, while the low yield trajectory (OSKM) appears to meander through inefficient directions. Thus the reprogramming dynamics of B cells are similar to  the reprogramming dynamics of MEF: in all cases reprogramming follows a simple one-dimensional reaction coordinate and can be understood as a  barrier crossing between minima. 

The insets in these figures show results from numerical simulations using the landscape model. The simulations reveal a simple reaction coordinate. However unlike in experiment, the simulated trajectories for the two protocols exhibit nearly identical dynamics. This likely reflects the limitations of the coarse-graining approximation used to construct the landscape model. In the model, TFs are treated as binary variables and all TFs are treated on equal footing -- no distinction is made between more promiscuous TFs like CEBP$\alpha$ and more specific downstream factors. Despite these limitations, the phenomenological model still captures the qualitative phenomena seen in the experiments.  

The similarity of the reprogramming trajectories from MEFs and B cells suggest a universal reaction coordinate for reprogramming: a straight line connecting the starting and ending cell type in projection space. This can be seen best in Figures 2C and 2D where we have plotted reprogramming dynamics from both MEFs and B Cells on the same plots. These experimental data are consistent with numerical simulations using our landscape model which show that reprogramming trajectories always follow a straight line in projection space for both choices of starting cell type. 

\subsection{Insight into Dynamics from Our Mathematical Model} 
Given the strong agreement between experiment and the landscape model, it is interesting to ask if the model can provide further insights into  reprogramming dynamics beyond those that can be directly gleaned from analyzing experimental time series. As discussed in the introduction, there is an ongoing debate in the reprogramming literature about the order and organization of gene-level events during reprogramming \cite{Buganim2012Single-Cell, Polo2012A-Molecular, Chung2014Single}. To address this, we performed detailed  simulations that allowed us to probe gene-level events during  reprogramming from MEF to iPSC (see Materials and Methods).  Experimentally, reprogramming times (as measured by reporters for pluripotency markers) are well described as a Poisson process, implying the existence of a single rate limiting step \cite{Hanna2009Direct}. Our simulation results support the idea of a single rate limiting step to the turning on of pluripotency markers (see Fig 3A). In our simulations, the time to turn-on pluripotency markers is calculated by measuring the time it takes a trajectory to have a significant projection on an iPSC state ($a^{end}=0.3$) Additionally, our simulations show that the later phase of reprogramming (defined as the period of time when trajectories go from having a projection $a^{end}=0.3$ to $a^{end}=0.8$) follows a narrowly peaked distribution. Once reprogramming has started, it is very fast: the median time for the later phase is approximately 40 times shorter than the median time for the early phase. Consistent with experiment \cite{Hanna2009Direct,Rais2013Deterministic}, we find that almost all trajectories eventually reprogram. In agreement with Yamanaka\cite{Yamanaka2009Elite}, these results are inconsistent with an ``elite'' model of reprogramming in which only a special subset of cells are amenable to reprogramming. 

To ask about the order of gene level events, we probed the gene level dynamics of 10 genes known to be mutually exclusive for either MEFs or ESCs (\emph{Snai1}, \emph{Snai2}, \emph{Prrx1}, \emph{Twist2}, \emph{Twist1} and  \emph{Zfp42}, \emph{Nanog}, \emph{Utf1}, \emph{Lin28a}, \emph{Sall4}, respectively) for 224 successful reprogramming trajectories out of a total of 3000 attempts. Recall, that in our model, each gene is represented by a binary variable and can either be `on' or `off'. Since the dynamics of our landscape model are stochastic, these genes turn on and off at different values of reaction coordinate in each of these 224 trajectories.  To understand if there is any structure in the gene level dynamics, we  counted the percentage of trajectories for which a gene was on on at a given reaction coordinate using a moving average (see Materials and Methods). The results are shown in Fig 3B (see SI Figure 10 for an example of non-averaged data). The MEF (ESC) genes gradually turn off (on) over time as expected. Furthermore, the order in which genes turn on and off is relatively stable, at least when averaged over trajectories.  In contrast, individual simulation trajectories show much more variability in the order which genes turn on. However, if we consider individual pairs of TFs, we find that their ordering tends to be consistent with what one would expect from Fig 3B. For example, \emph{Nanog} turns on before \emph{Sall4} in 58\% of trajectories, and \emph{Snai1} turns off before \emph{Twist1} in 71\% of trajectories, but for \emph{Twist1} and \emph{Twist2}, there is no clear trend for one or the other to turn off first. 

All the qualitative features of our simulations are consistent with the idea that reprogramming trajectories correspond to successful ``barrier crossing'' between two minima in a landscape. An important qualitative prediction of all barrier crossing is that reprogramming trajectories should be dominated by a small number of optimal paths, with some amount of fluctuations around those paths\cite{Roma2005Optimal,Mehta2008Exponential}. In particular, the facts that the early phase of reprogramming is well described by a Poisson process, the later phase is described by a narrow distribution of times, and that the median time for the early phase is much longer than the median time for the later phase are all features that would be expected of a simple barrier-crossing process. Furthermore, these simulations show that in a high-dimensional barrier crossing, genes can turn on in a temporally ordered manner (at least when averages over many reprogramming attempts) even though the process is driven entirely by stochasticity.

\section{Discussion}
A common metaphor used to describe cellular identity is Waddington's landscape or the idea of a rugged ``epigenetic landscape'' in which cell types are basins of attraction. In this picture, cellular reprogramming is envisioned as a process in which one cell type is externally driven out of its basin of attraction, across a barrier, and eventually ends up in the basin of attraction of the desired cell type. Previously, we used ideas from spin physics to introduce a model of cellular identity that can be built from genome expression data. In this paper, we reanalyzed experimental data on reprogramming dynamics in terms of our model and found good agreement between the experiments and simulations of our model. 

Our model provides several interesting insights into reprogramming dynamics.We find that reprogramming dynamics proceed along a simple one-dimensional reaction coordinate and must cross a significant energy barrier. Somewhat surprisingly, this reaction coordinate is independent of reprogramming dynamics. In terms of projections, we can simply describe the reaction coordinate as a straight line from ($a^{start}=1$, $a^{end}=0$) to ($a^{start}=0$, $a^{end}=1$). What makes this simple picture especially interesting is that we demonstrated its validity for two different types of reprogramming experiments (MEF or B Cell to iPSC). Based on simulations with our model, we believe that any cellular interconversion (reprogramming or direct conversion), will proceed along a similar, universal, reaction coordinate when described in terms of $a^{start}$, $a^{end}$, and energy.  We expect this basic picture should also be valid in other organisms such as humans.

Our model also gives insight into the ongoing debate about the phases of reprogramming dynamics. A priori, reprogramming dynamics may be either probabilistic or deterministic with respect to both the timing and order of gene level events. Our simulations show the the initial phase of reprogramming follows a Poisson distribution -- initiating reprogramming is a rare event. However, once initiated, reprogramming proceeds quickly and efficiently. This is reflected in our simulations by the observation that the dynamics of the reprogramming process at later stages are well described by a narrowly peaked distribution. Furthermore, we find that when averaged over many successful reprogramming trajectories, the order of gene level events are relatively reproducible. Our simulations strongly support Chung et al\cite{Chung2014Single} description of reprogramming as a ``loosely ordered probabilistic process''. 

Why have different dynamics experiments led to such drastically different conclusions? So far, each experiment has used different techniques, each of which have their own limitations. GFP reporters (for example \cite{Hanna2009Direct}) provide precise timing data but are limited to small numbers of genes. Whole genome expression data (for example \cite{Polo2012A-Molecular}) provides data on all genes, but both microarrays and RNA-Seq require populations of cells. Finally, single cell gene expression data (for example \cite{Buganim2012Single-Cell}) provides accurate details of gene expression, but only for a subset of genes (currently 48 with standard Fluidigm chips\cite{Buganim2012Single-Cell}). Therefore, depending on which technique is utilized, each experimentalist rightfully sees a different picture of reprogramming dynamics. However, viewing reprogramming as a loosely ordered probabilistic process unifies all of these different experimental pictures. 

Besides examining the gene level reprogramming dynamics, our model provides a clearer picture of the global mechanism behind reprogramming. One of the most surprising aspects of reprogramming is that the  over expression of just a few TFs (out of thousands) can lead to such drastic changes in the global gene expression profile. Our simulations suggest the underlying reason for this is the important role played by culturing conditions. In our model, inducing the OSKM TFs in MEFs only changes the energy by 0.5\% , which at the noise levels considered here, do not lead to any successful reprogramming event. However, by including the effect of cell culture in our simulations, we achieve 7.43\% reprogramming rates. This suggests that culturing conditions likely play an important role in dictating reprogramming efficiencies. For example, it is claimed\cite{Kim2011Direct} that it is possible to use the OSKM factors, normally used to reprogram to iPSC, to instead reprogram to neuronal progenitors just by changing culture conditions. This highlights an important issue of experimental design for direct conversions to a given cell type. Before one searches for TFs to manipulate, it is essential to understand the correct culturing conditions for the desired cell type. Without the correct medium, direct conversion may prove exceedingly difficult. In our simulations, we have found that the culture term for a given cell type decreases the size of the basin of attraction of all the other cell types. We even find some reprogramming events when we bias the system just by introducing the culture term, without forcing expression of the OSKM TFs (this likely reflects the limitations of the model). However, when we compare simulations of MEF to ESC reprogramming at a certain noise level and for a certain duration, the ones where expression of the OSKM TFs is forced and the ESC culture term is present have a success rate 5 times higher than the ones where the ESC culture term is present but OSKM expression is not forced. In the future, it will be interesting to further explore this tradeoff between stability and plasticity of cell types.

The experimental analysis and simulations presented here suggest that reprogramming can be viewed as a ``barrier crossing'' in rugged landscape (see Figure 4). In all barrier crossings, the dynamics are dominated by a few ``optimal paths'', suggesting that reprogramming dynamics are likely to be low-dimensional and fairly reproducible at the gene level. A natural consequence of this picture is the existence of a simple reaction coordinate that describes the progress along the optimal path. If the landscape picture is correct, the existence  of a reaction coordinate is likely to be a generic feature of all reprogramming and direction conversion protocols. Directed differentiation is a closely related experimental technique that instead of using TFs to convert between cell types focuses on recapitulating embryonic development through sequences of signaling molecules\cite{Wilson2015Emergence}. It will be interesting to see if projections are also a useful reaction coordinate for directed differentiation experiments.

The results presented here are also likely to be applicable to other systems. Recently, it has been suggested the evolutionary dynamics of viruses such as HIV can also be understood using a Hopfield-inspired landscape model\cite{Ferguson2013Translating}. In evolutionary landscapes, crossing fitness valleys in rugged landscapes can naturally be understood in terms of barrier crossings. For this reason, it is likely that the techniques developed here in the context of cellular reprogramming can be adapted to visualize evolutionary data on fitness crossing dynamics. More generally, landscapes have proven to be an important tool for furthering our understanding a variety of biological problems such a protein folding \cite{Bryngelson1995Funnels,Onuchic1997Theory,Onuchic2004Theory}. The intuitions developed in the context of these other problems are also likely to be applicable to cellular reprogramming and in the future, it will be interesting to explore these connections further.



\section{Materials and Methods}
\subsection{Data Analysis}
Here we present details of the data analysis. All the experimental data used in this paper are available in the online Supplementary Files. 
\begin{itemize}
\item Metadata-Cell\_Type\_basis.txt. List of publicly available microarrays used to define cell types. 
\item Metadata-Data\_Analysis.txt. Data samples analyzed in this paper.
\item Cell\_Type\_Basis-Data\_Analysis.txt. Zscore data that defines cell types for the data analysis.
\item Cell\_Type\_Basis-Simulations.txt. Binarized data that defines cell types for the simulations.
\item Data\_Analysis.txt. Zscore data for the data samples.
\end{itemize}

The following abbreviations are used in the Supplementary Files. All GSE and GSM are data identifiers from \href{http://www.ncbi.nlm.nih.gov/geo}{\textcolor{cyan}{NCBI GEO}} except that GSE labels E-MEXP are data identifiers from \href{http://www.ebi.ac.uk/arrayexpress/}{\textcolor{cyan}{ArrayExpress}}. Sample\_Name refers to the sample label in Data\_Analysis.txt. Plot\_Label refers to the abbreviations used in Figures 1 and 2. Paper\_Reference is an abbreviated citation of the source data which came from the following papers\cite{Sridharan2009Role,Mikkelsen2008Dissecting,Rais2013Deterministic,Polo2012A-Molecular,Samavarchi-Tehrani2010Functional,Koga2014Foxd1,Fluri2012Derivation,Di-Stefano2014C/EBPalpha,Di-Stefano2014Time-resolved}.  

Microarrays were taken from public datasets and come from a variety of different microarray platforms. In order to compare the different platforms, the following analysis was done. The raw microarray data was converted to an expression level as follows. Microarray probe-to-gene map was created with Bioconductor 3.0. All raw microarray files were initially processed by robust mean averaging (RMA) and genes with multiple microarray probes were averaged. Since we were interested in cellular identity, only transcription factors, transcription co-factors, or chromatin remodeling genes were kept (for short hand, referred to as transcription factors,TF, throughout the text)\cite{Zhang2014AnimalTFDB}. 

While the above analysis was done for both experimental data and simulations, from this point on the analysis differed between the two cases. For the experimental data analysis, we only used TFs that were common to all of the different microarray platforms, leaving $N=994$ TFs. In order to make robust comparisons across platforms the RMA output was converted to a rank order. Next, we wanted to convert this rank order to the z-score of a log-normal distribution. We converted the rank to a percentile (for N genes, by dividing by N+1), and then this percentile into a normal z-score. For later mathematical convenience, we used a biased estimator (i.e. we normalized by N and not N-1) since then the Euclidean norm of each microarray gene expression was N. Therefore, for the data analysis each sample is described by a Gaussian distribution with a Euclidean norm of $N=994$.

For the simulations, we followed similar steps to produce continuous TF expression levels for the cell type basis vector. However, in order to reduce the computational cost, we binarized the gene expression so that each TF is either on $(+1)$ or off $(-1)$. We then dropped all TFs that were always on or always off in every cell type, leaving $N=1436$ TFs for the simulations.

\subsection{Cellular Identity Landscape}
Here we summarize our model for the cellular identity landscape\cite{Lang2014Epigenetic}. The $N$ transcription factors (TF) are labeled by Latin indices $i$ and the $p$ cell types are labeled by Greek indices $\mu$. When analyzing experiments, we keep the $N=994$ TFs common to all of the experimental datasets.  Each sample is a Gaussian distribution with mean equal to $0$ and Euclidean norm equal to $N$. This implies a standard deviation of $\frac{N}{N-1}\approx 1$. When performing simulations, we use the complete set of $N=1436$ TFs and each TF is either on $(+1)$ or off $(-1)$. A general network state is represented by a vector $S_i$ of length $N$. A cell type $\mu$ is represented by the vector $\xi_i^\mu$. The correlation (dot-product, overlap, or magnetization) between an arbitrary state and the cell type $\mu$ is given by
\begin{equation}
m^\mu=\frac{1}{N}\displaystyle\sum_{i=1}^N \xi_i^\mu S_i.
\end{equation}
The correlation matrix between cell types is given by
\begin{equation}
A^{\mu\nu}=\frac{1}{N}\displaystyle\sum_{i=1}^N \xi_i^\mu \xi_i^\nu .
\end{equation}
The projection onto each cell type is
\begin{equation}
a^\mu=\displaystyle\sum_{\nu=1}^p (A^{-1})^{\mu\nu}m^\nu .
\end{equation}
We require all $\xi_i^\mu$ to be attractors in the landscape. This is ensured by constructing a correlation-based interaction network given by
\begin{equation}
J_{ij}=\frac{1}{N}\displaystyle\sum_{\mu=1}^{p} \displaystyle\sum_{\nu=1}^{p} \xi_i^\mu (A^{-1})^{\mu\nu} \xi_j^\nu
\end{equation}
with $A^{\mu\nu}$ the correlation matrix between cell types. This interaction network produces stable basins of attraction and is written in terms of the cellular identity landscape as
\begin{equation}
H_{\text{basin}} = -\frac{1}{2}\sum_{ij=1}^N S_i J_{ij} S_j = -\frac{N}{2}\sum_{\mu=1}^p m^\mu a^\mu, 
\end{equation}
which is a Lyapunov function. 
There is also a culture term 
\begin{equation}
H_{\text{culture}} = - \sum_{\mu=1}^p b^\mu a^\mu, 
\end{equation}
which stabilizes specific cell types. Normally only one cell type $\mu$ is stabilized, by choosing $b^{\mu} > 0$, and all the other coefficients $b^{\mu'}$ are zero. 

The complete landscape is defined by an abstract energy $H$, which is composed of the two terms just discussed:
\begin{eqnarray}
H &=& H_{\text{basin}} + H_{\text{culture}}  \nonumber\\
 &=& -\frac{N}{2}\sum_{\mu=1}^p m^\mu a^\mu  - \sum_{\mu=1}^p b^\mu a^\mu 
\end{eqnarray}
In addition, the model allows for certain transicription factors $i$ to be locked, typically by choosing them to be ``on'', i.e.~$S_i = +1$.

\subsection{Dynamics and Simulations}

To numerically study the dynamics of the model, we assume that the TFs can probabilistically switch states. To save computational effort, we also assume that the expression values are binarized, i.e.~the only possible expression values are $+1$ and $-1$. Each TF is biased towards a state by its interactions with the network through its local field $h_i=-\frac{\partial H}{\partial S_i}$. The evolution is probabilistic and controlled by a global noise parameter $\beta$ (i.e.~ inverse temperature $\beta=1/T$). At each simulation update, $u$, one randomly chosen TF $i$ is updated. The probability of the value $S_i$ at update $u+1$ is related to the local field $h_i(u)$ at update $u$ by
\begin{equation}
P[S_i,u+1]=\frac{e^{\beta h_i(u)S_i}}{e^{\beta h_i(u)} + e^{-\beta h_i(u)}}.
\label{eq:heat-bath}
\end{equation}
Additionally, as indicated above, in some of the simulations a subset of the transcription factors is locked at a certain value. Concretely, in many of the simulations we discuss, the OSKM TFs are fixed to have the value $+1$ (``on'').  

We performed Monte Carlo (MC) simulations of a system containing $N=1436$ TFs using the update rule given by Eq.~(\ref{eq:heat-bath}), with noise parameter $\beta = 1.62$ (i.e.~ $T\approx 0.617$). When a culture term was introduced, it was to bias the system towards the ESC cell type, with $b^{\mu} = 0.03$ for $\mu = \text{ESC}$ and $b^{\mu'} = 0$ for all other cell types. The dynamics are qualitatively similar for a wide-range of values for $b^{\mu}$ and the quantitative differences will be explored in future work.

Most of the results reported in this paper correspond to simulations where the total number of steps was $t=10^5$. For the case of simulations of MEF to ESC reprogramming, the OSKM transcription factors were locked ``on'' for the whole simulation, and the culture term was present from step $t=5000$ until the end. In this case, 3000 trajectories were simulated, out of which 224 successfully reprogrammed, i.e.~the reprogramming rate was  7.43\%. For the simulations of B-cell to ESC reprogramming, the protocol was similar, and in this case 205 trajectories reprogrammed successfully out of a total of 3000, corresponding to a reprogramming rate of 6.83\%. 

In order to obtain additional details about the probability distributions of times associated with the reprogramming, which we show in Figure 3A and SI Figure 11, we performed an additional set of simulations of MEF to ESC reprogramming, with the only change being that the total number of steps was 30 times larger, i.e.~$t = 3\times 10^6$ instead of $t = 10^5$. In this set of much longer simulations, 2937 trajectories out of 3000 successfully reprogrammed from MEF to ESC, which corresponds to a reprogramming rate of 97.90\%. 


\section{Acknowledgments}
We thank members of the Boston University Center for Regenerative Medicine (CReM) for extremely useful discussions. STP acknowledges the Condensed Matter and Surface Science (CMSS) program at Ohio University for support through a studentship. AHL was supported by a National Science Foundation Graduate Research Fellowship (NSF GRFP) under Grant No. DGE-1247312. PM was supported by a Simon's Investigator in the Mathematical Modeling in Living Systems. This work was supported in part by Ohio University. 

\bibliography{bibliography}

\newpage

\begin{figure}
\includegraphics[width=15.8cm]{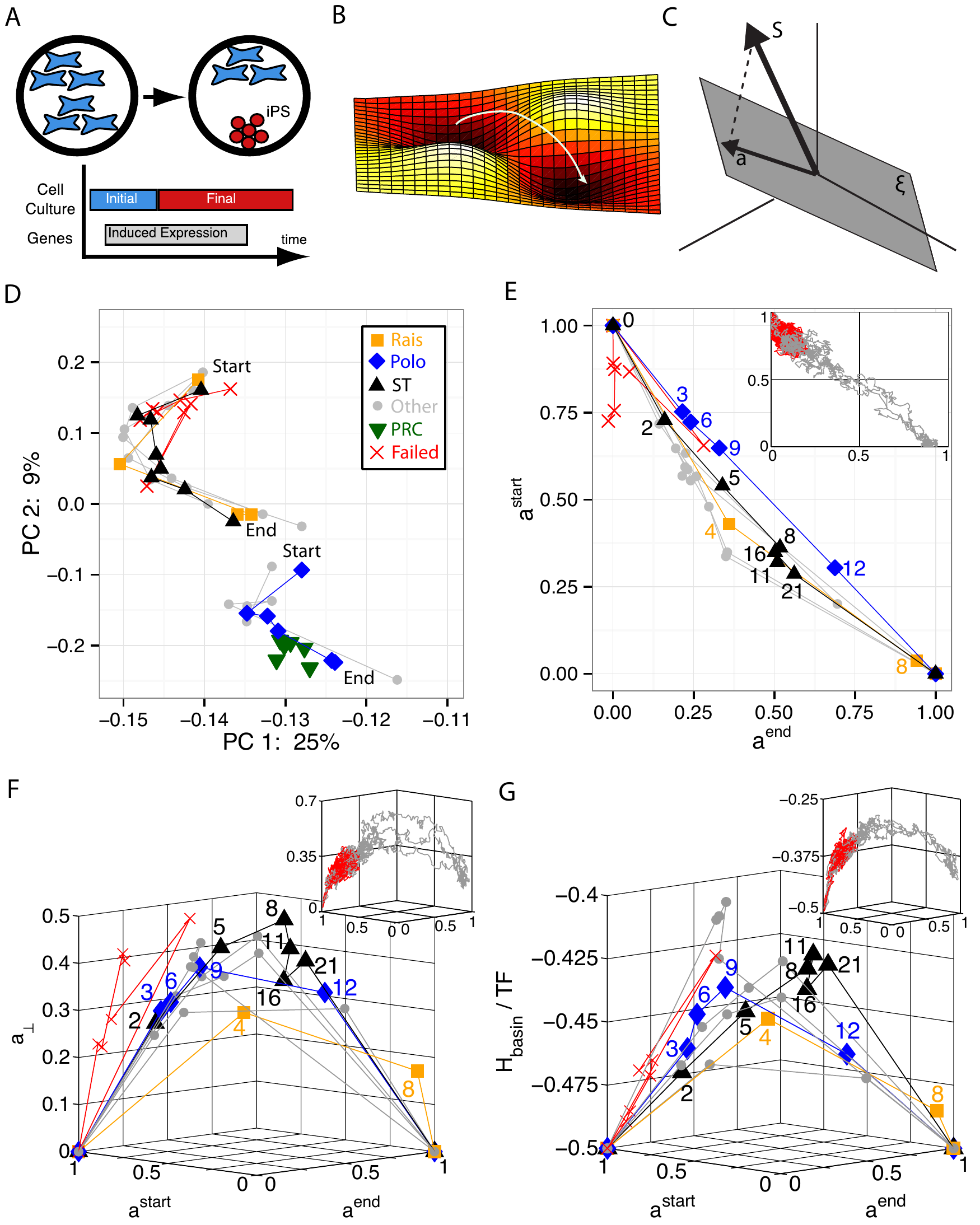}
\caption{\textbf{Cellular Reprogramming Reaction Coordinate.} A. Transient expression of reprogramming genes plus switching culturing conditions probabilistically leads to the desired cell type. B. Reprogramming is commonly described as the crossing of a barrier in a high-dimensional landscape. C. Our proposed cellular identity landscape is based on the projection, $a$, of an arbitrary gene expression, $S$, onto the subspace (gray plane) spanned by the natural cell types, $\xi$. D. Principal component analysis (PCA) of reprogramming from mouse embryonic fibroblasts (MEF) to induced pluripotent stem cells (iPSC) with start marking day 0 and end marking iPSC. Rais\cite{Rais2013Deterministic}, Polo\cite{Polo2012A-Molecular}, and ST (Samavarchi-Tehrani)\cite{Samavarchi-Tehrani2010Functional} are three successful trajectories in which the explicit time in days is labeled on plots E, F, and G. Other represents additional successful trajectories, PRC are partially reprogrammed cells, and failed trajectories do not reprogram. E. Projection onto $a^{start}$ (MEF) and $a^{end}$ (iPSC) only. All successful trajectories follow a simple reaction coordinate in projection space, a straight line from ($a^{start}=1$,  $a^{end}=0$) to ($a^{start}=0$, $a^{end}=1$). Insets in E, F, and G are simulation data with failed trajectories in red and successful trajectories in gray. See SI Fig 2 for larger version of simulations. F. Measure of projection on all other cell types, $a_{\perp}$ vs the reaction coordinate. See SI Fig 3 for larger version of simulations. G. Energy landscape of basins of attraction, $H_{\text{basin}}$, per transcription factor (TF) vs reaction coordinate. See SI Fig 4 for larger version of simulations.
}
\end{figure}

\begin{figure}
\includegraphics[width=15.8cm]{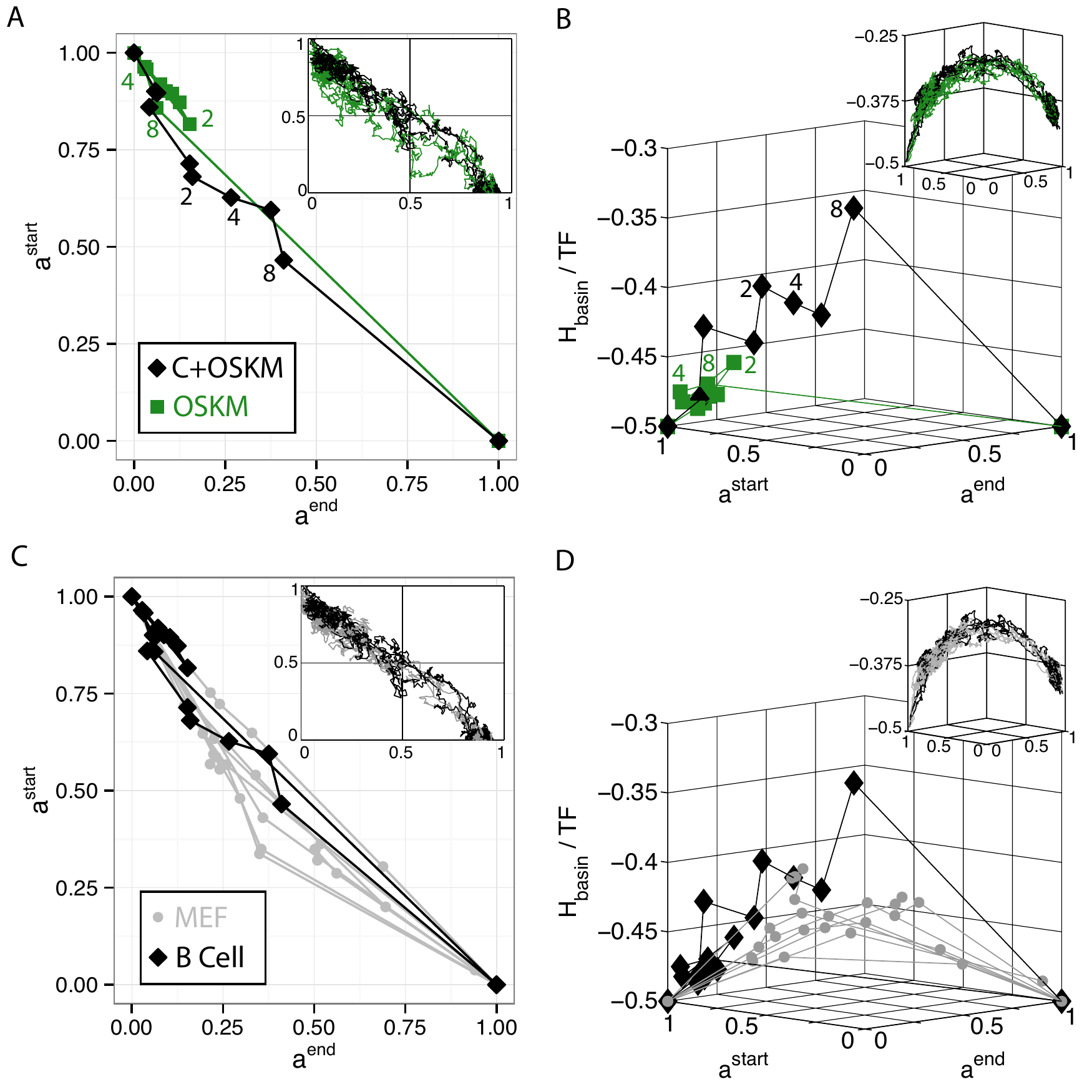}
\caption{\textbf{Universal Reaction Coordinate.} A. Cellular reprogramming from $a^{start}$ (B Cells) to $a^{end}$ (iPSC) by Di Stefano et al\cite{Di-Stefano2014C/EBPalpha}. OSKM is the standard Yamanaka protocol, while C+OSKM is a pulse of C/EBP$\alpha$ followed by OSKM which led to higher reprogramming yield. All insets are simulation data of same data shown in main figure. See SI Fig 6 for larger version of simulations.  B. Energy landscape of basins of attraction, $H_{\text{basin}}$, per transcription factor (TF) vs reaction coordinate. See SI Fig 7 for larger version of simulations. C. Data collapse of trajectories to $a^{start}$ vs $a^{end}$ for both MEF to iPSC (gray) and B Cell to iPSC (black). See SI Fig 8 for larger version of simulations. D. Data collapse of trajectories when viewed as energy vs reaction coordinate. See SI Fig 9 for larger version of simulations
}
\end{figure}

\begin{figure}
\includegraphics[width=15.8cm]{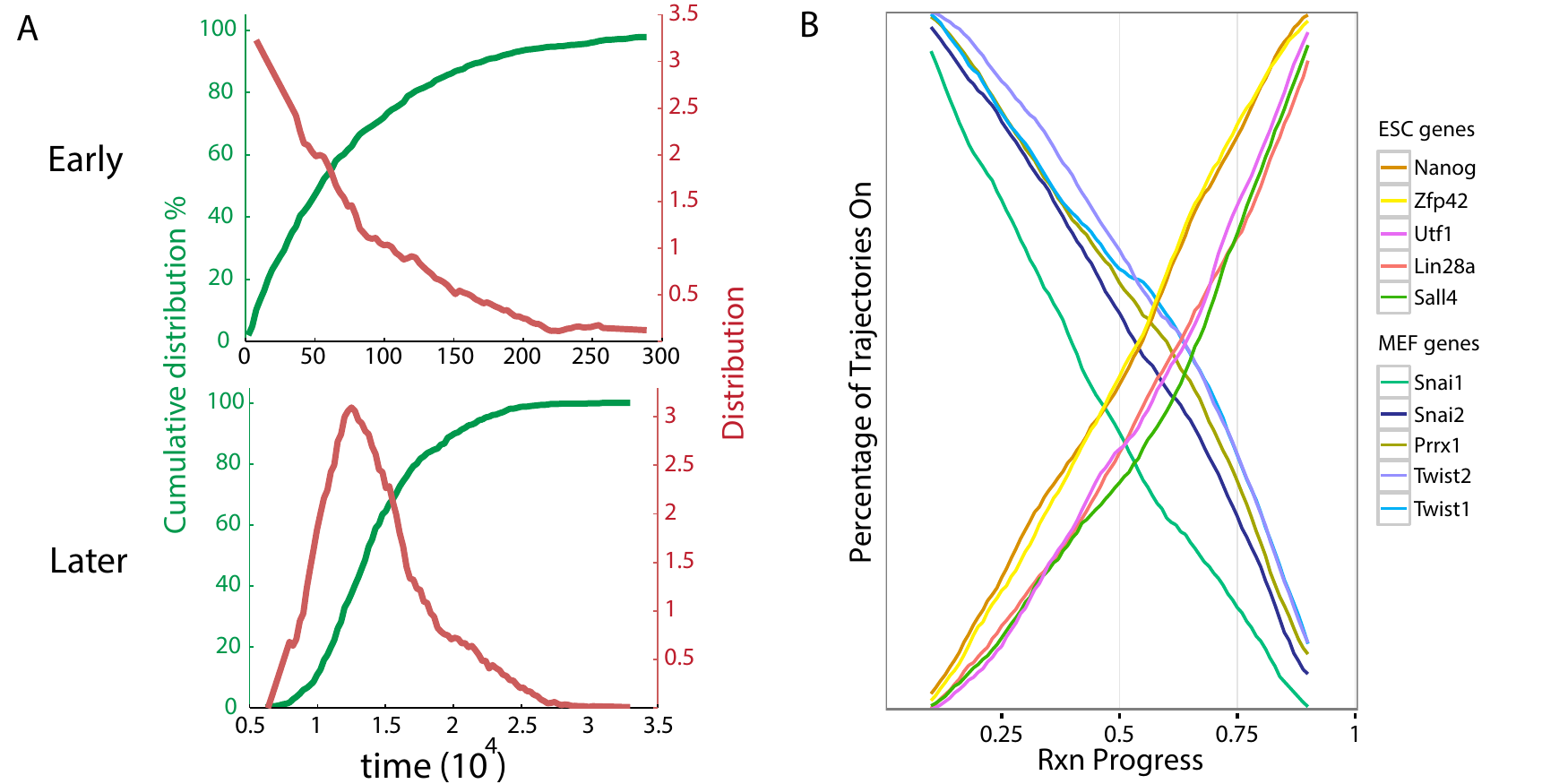}
\caption{\textbf{Nature of Reprogramming Dynamics in the Landscape Model.} A. Cumulative distributions of timing show that the early ($a^{end}=0$ to $a^{end}=0.3$) and later ($a^{end}=0.3$ to $a^{end}=0.8$) stages of reprogramming are respectively a Poisson and a narrowly peaked distribution. See SI Figure 11 for early ($a^{end}=0$ to $a^{end}=0.3$), middle ($a^{end}=0.3$ to $a^{end}=0.7$) and late ($a^{end}=0.7$ to $a^{end}=0.8$) phases of reprogramming as Poisson, narrowly peaked and narrowly peaked distributions respectively. In order to study the complete timing distribution, the data shown here and in SI Figure 11 were obtained in a simulation of duration $t=3 \times 10^6$ MC steps, which is $30$ times longer than the simulations reported on in all other figures. B. Percentage of trajectories in which a gene is on vs reaction coordinate. Data shown is a moving average of MEF (ESC) genes turning off (on) over time. See SI Figure 10 for example of non-averaged data.}
\end{figure}

\begin{figure}
\includegraphics[width=8cm]{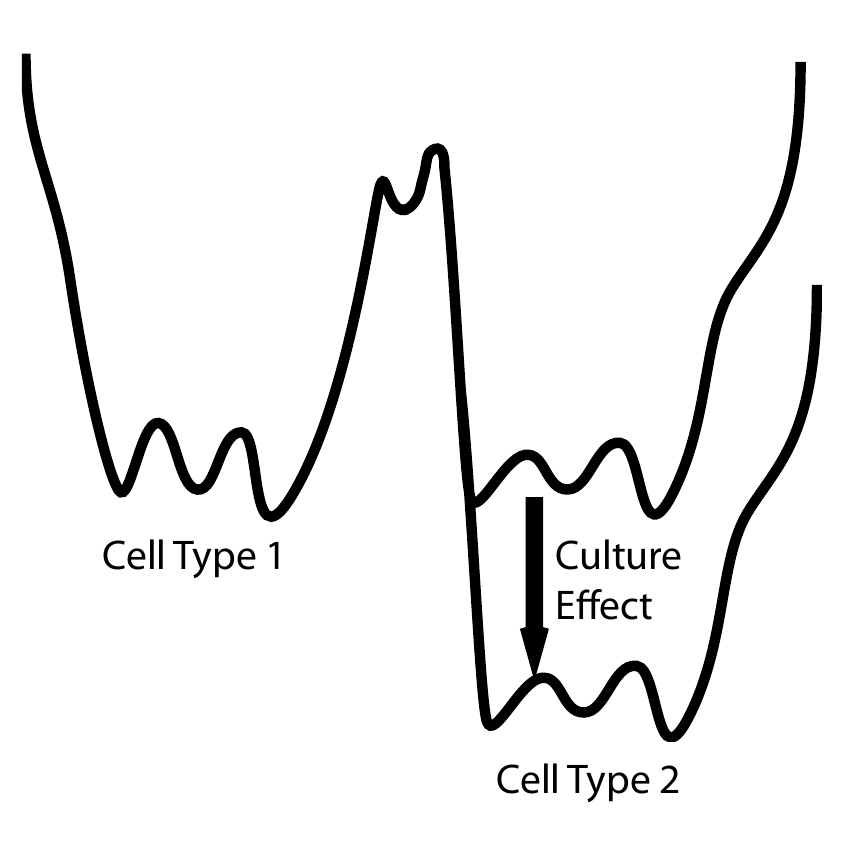}
\caption{\textbf{Culture Schematic.}
The correct culture conditions plays an essential role in reprogramming by stabilizing the final cell type.}
\end{figure}

\begin{figure}
\includegraphics[width=10cm]{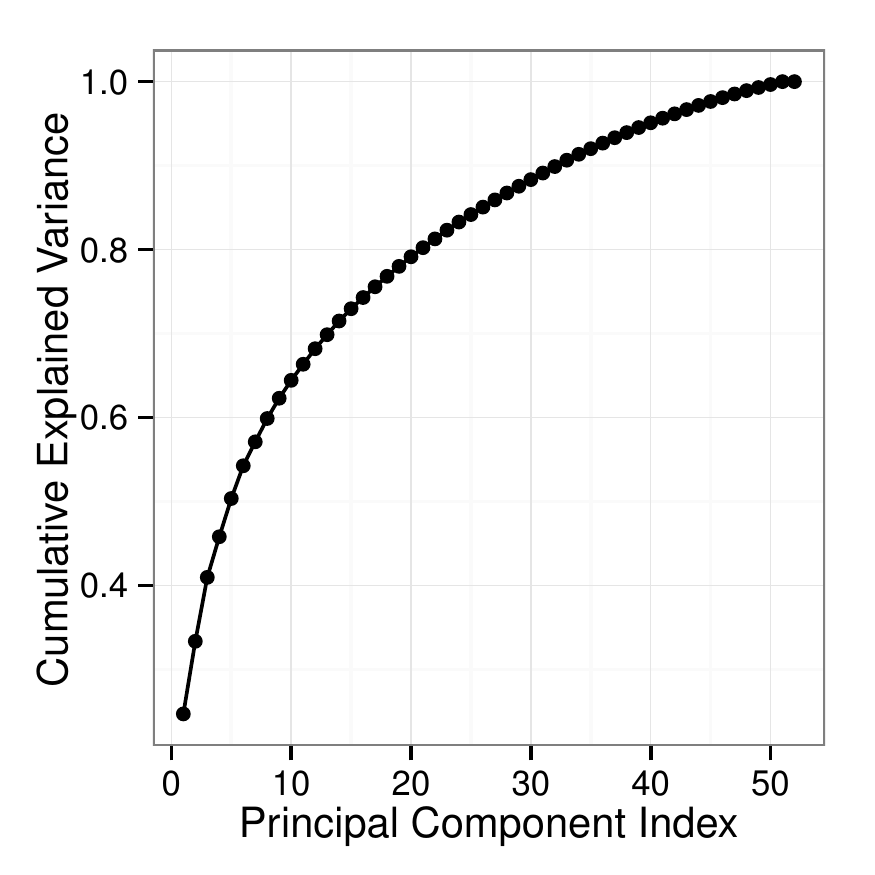}
\caption{\textbf{SI Figure 1. Principal components and explained variance.} This plot provides extended details of the principal component analysis (PCA) in Figure 1D. The cumulative fraction of explained variance vs principal component shows that in terms of PCA, the reprogramming dataset is high dimensional.}
\end{figure}

\begin{figure}
\includegraphics[width=10cm]{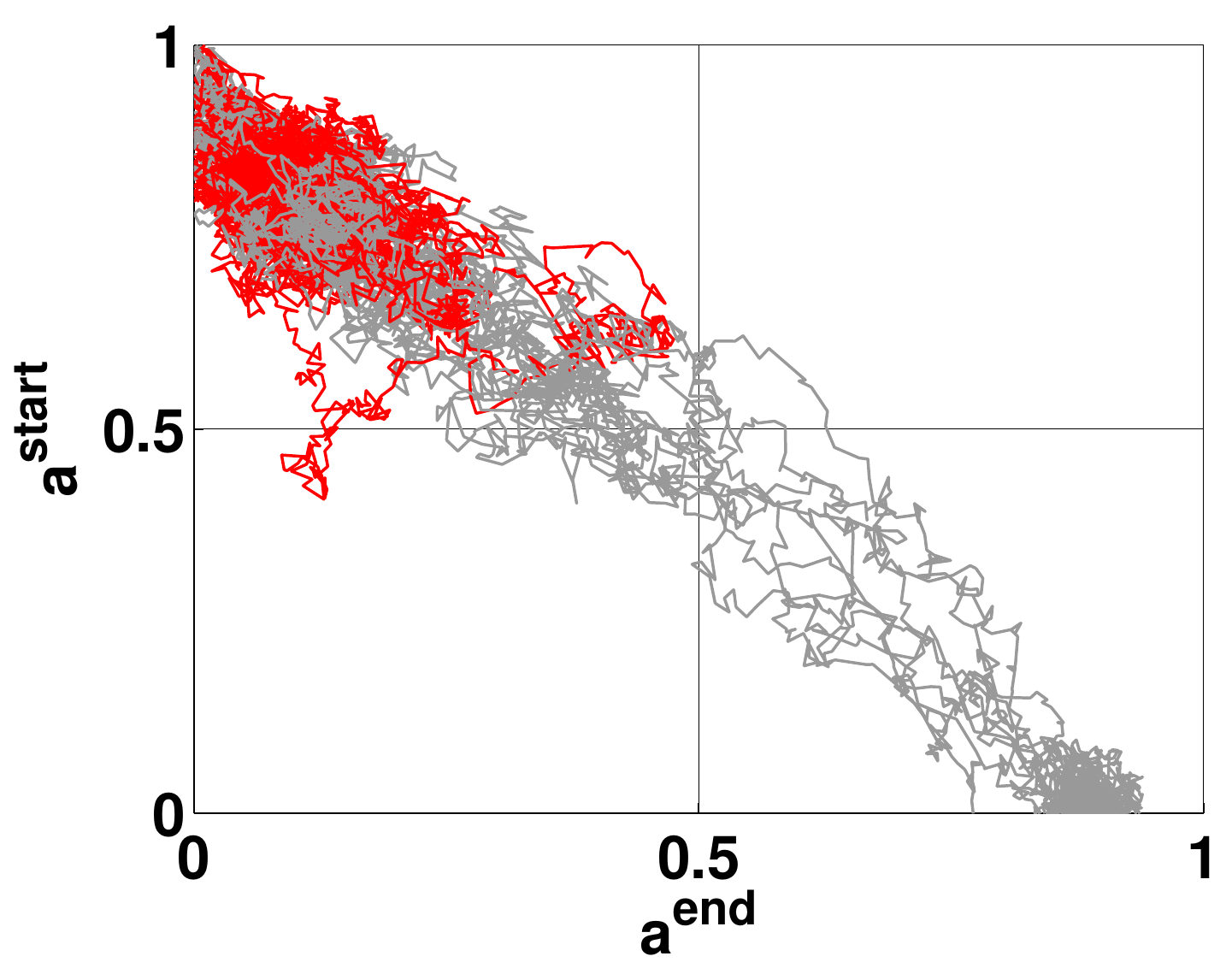}
\caption{\textbf{SI Figure 2. Fig 1E simulations.} Figure 1E simulation inset enlarged and with more trajectories.}
\end{figure}

\begin{figure}
\includegraphics[width=10cm]{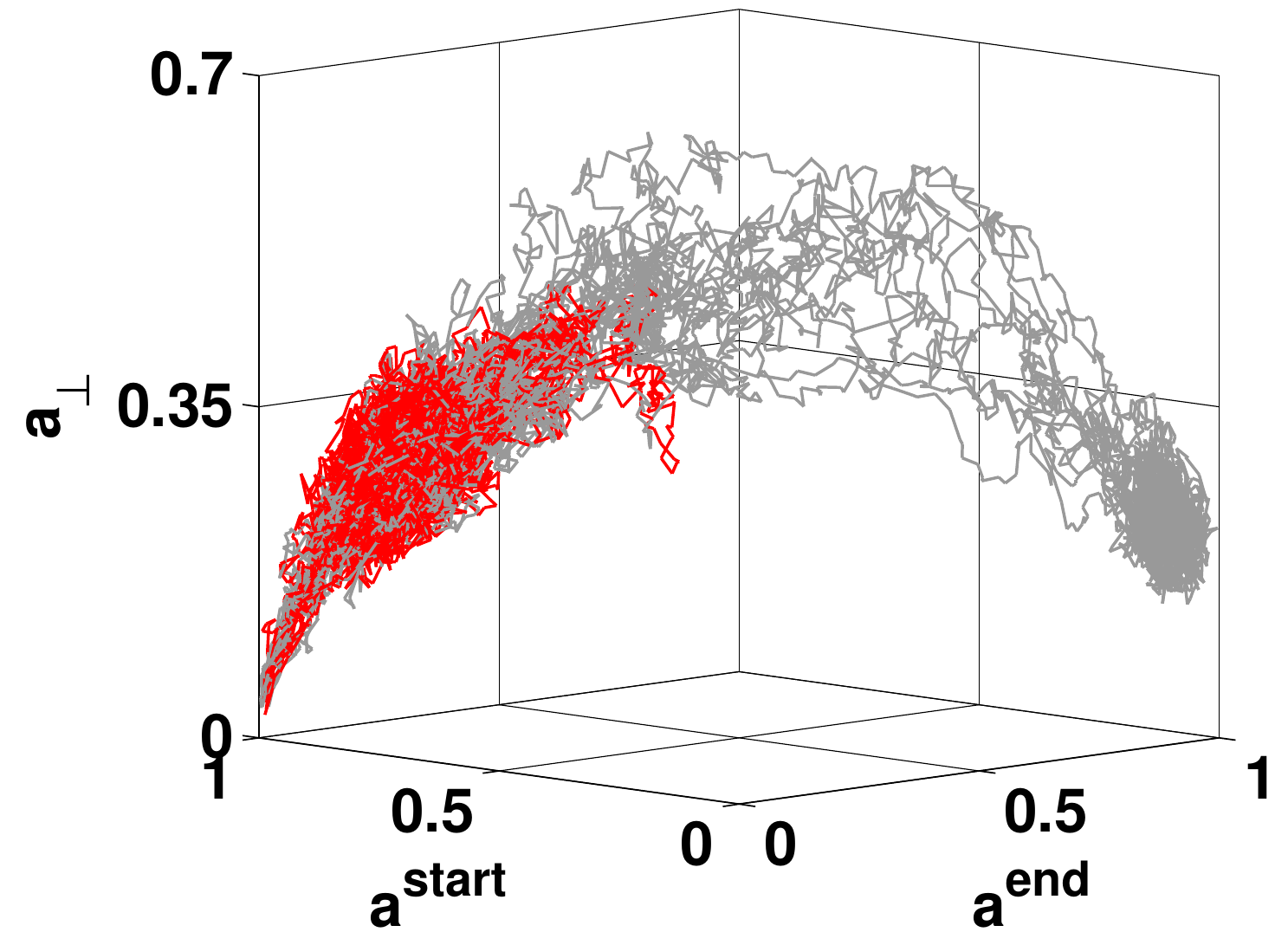}
\caption{\textbf{SI Figure 3. Fig 1F simulations.} Figure 1F simulation inset enlarged and with more trajectories.}\end{figure}

\begin{figure}
\includegraphics[width=15.8cm]{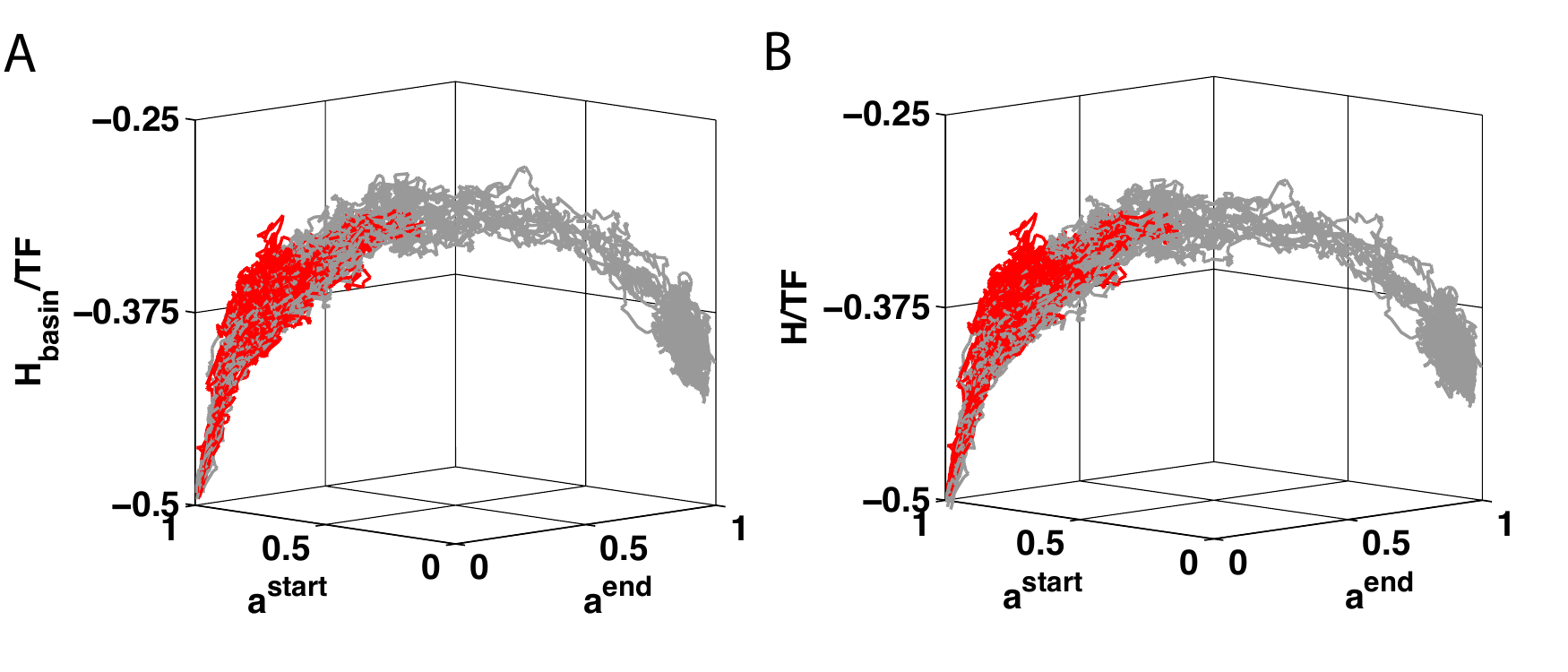}
\caption{\textbf{SI Figure 4. Fig 1G simulations.} A. Figure 1G simulation inset enlarged and with more trajectories. B. Figure 1G simulations including the small correction due to the culture term $H_{\text{culture}}$.}
\end{figure}

\begin{figure}
\includegraphics[width=15.8cm]{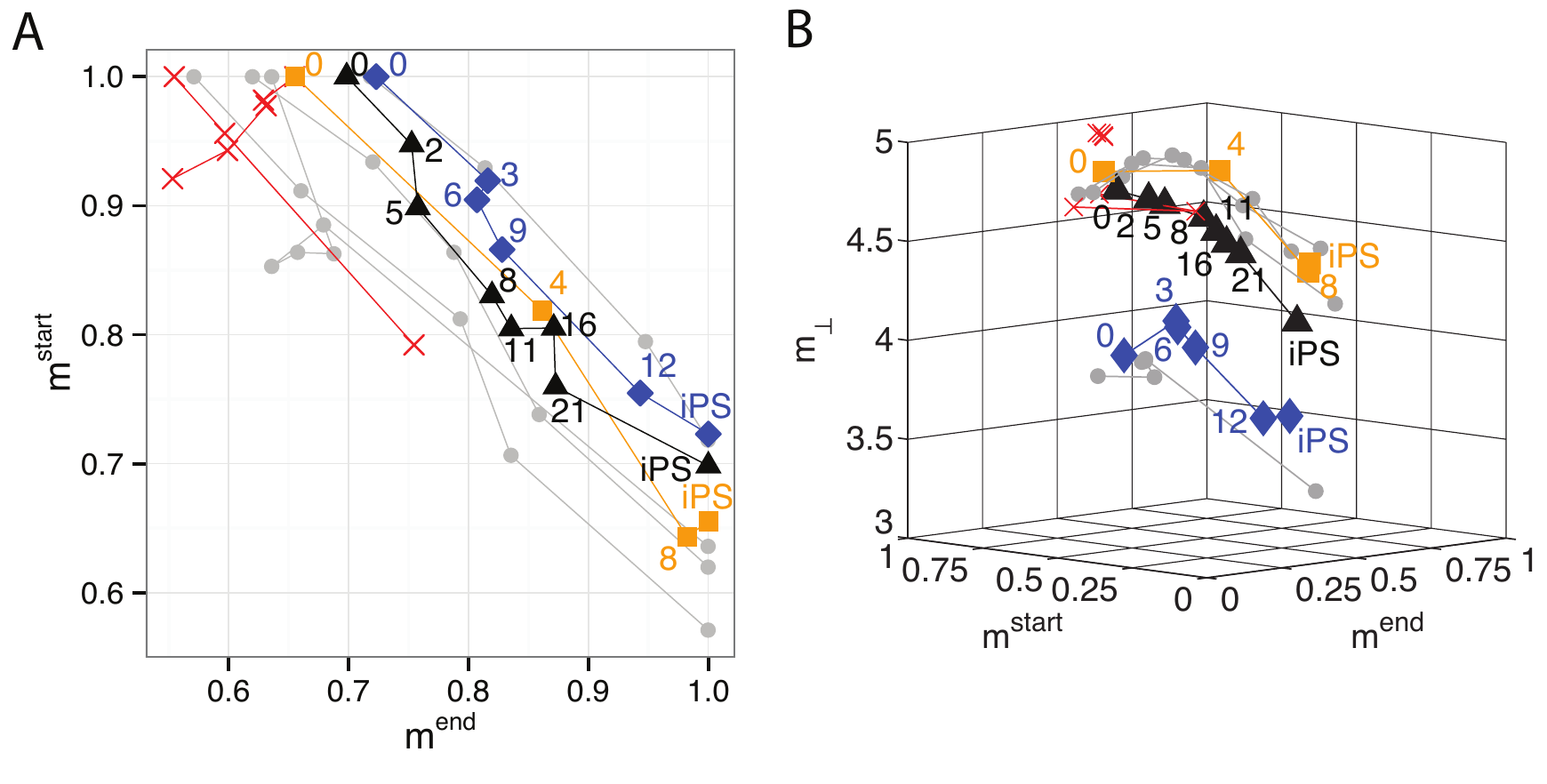}
\caption{\textbf{SI Figure 5. Alternative reaction coordinate and barrier.} A. This figure shows the same data presented in Fig 1E but instead of using projections ($a$), we have plotted dot products ($m$). B. This figure shows the same data presented in Fig 1F but instead of using projections ($a$), we have plotted dot products ($m$).}
\end{figure}

\begin{figure}
\includegraphics[width=10cm]{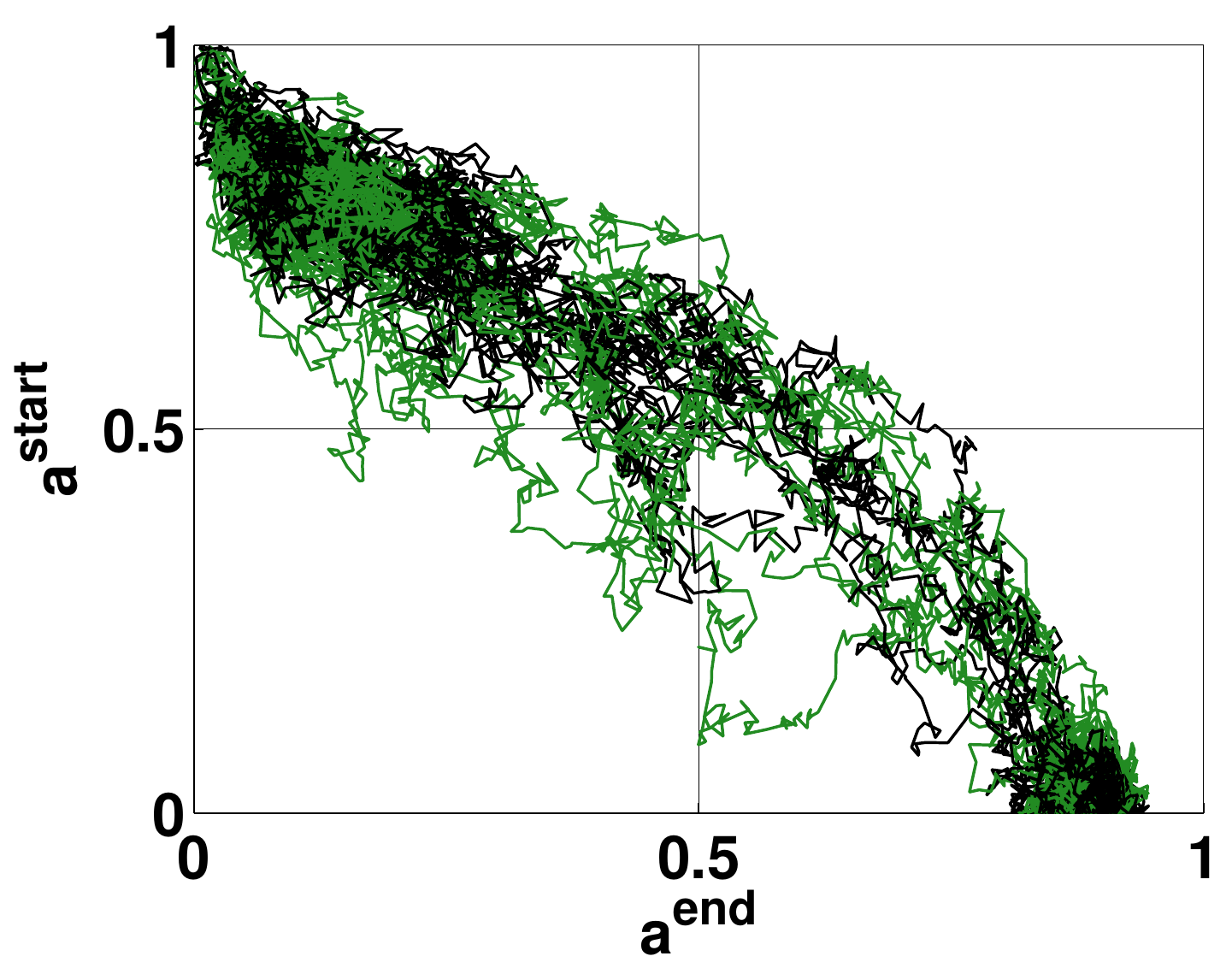}
\caption{\textbf{SI Figure 6. Fig 2A simulations.} Figure 2A simulation inset enlarged and with more trajectories.}\end{figure}

\begin{figure}
\includegraphics[width=15.8cm]{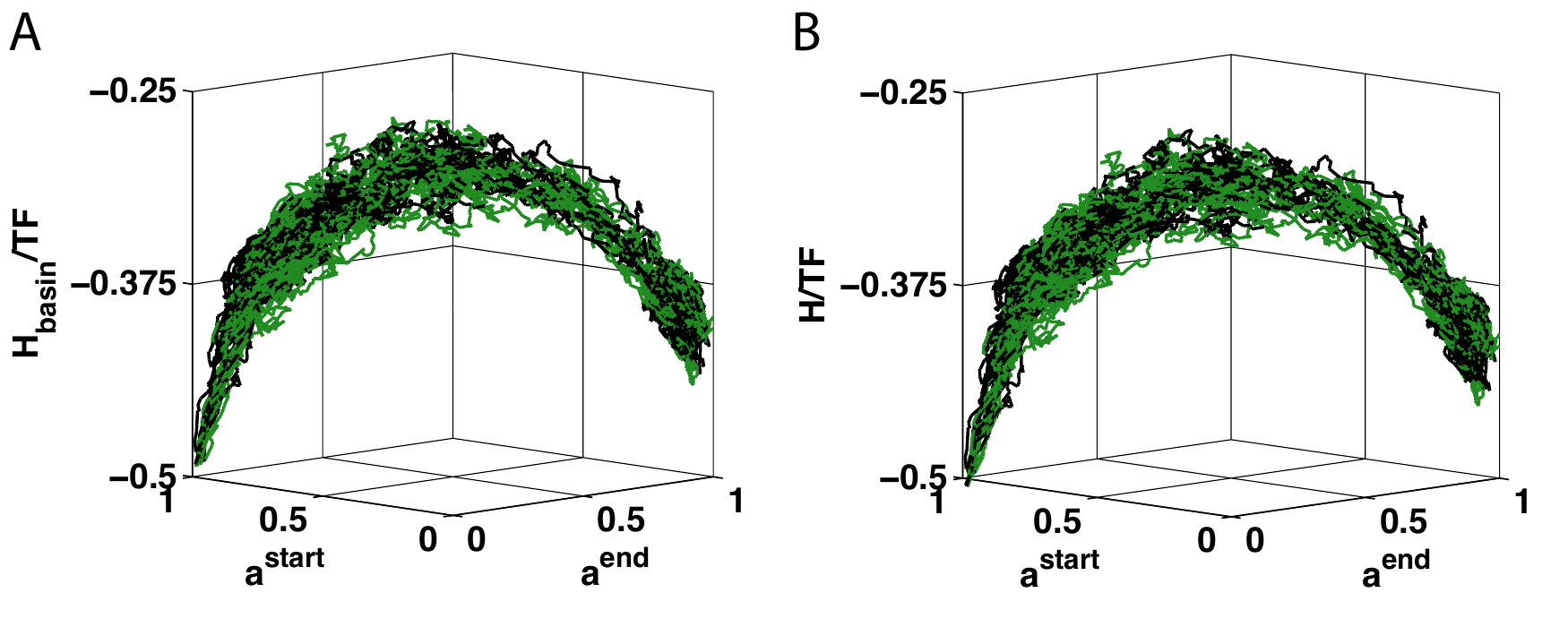}
\caption{\textbf{SI Figure 7. Fig 2B simulations.} A. Figure 2B simulation inset enlarged and with more trajectories.  B. Figure 2B simulations including the small correction due to the culture term $H_{\text{culture}}$.}
\end{figure}

\begin{figure}
\includegraphics[width=10cm]{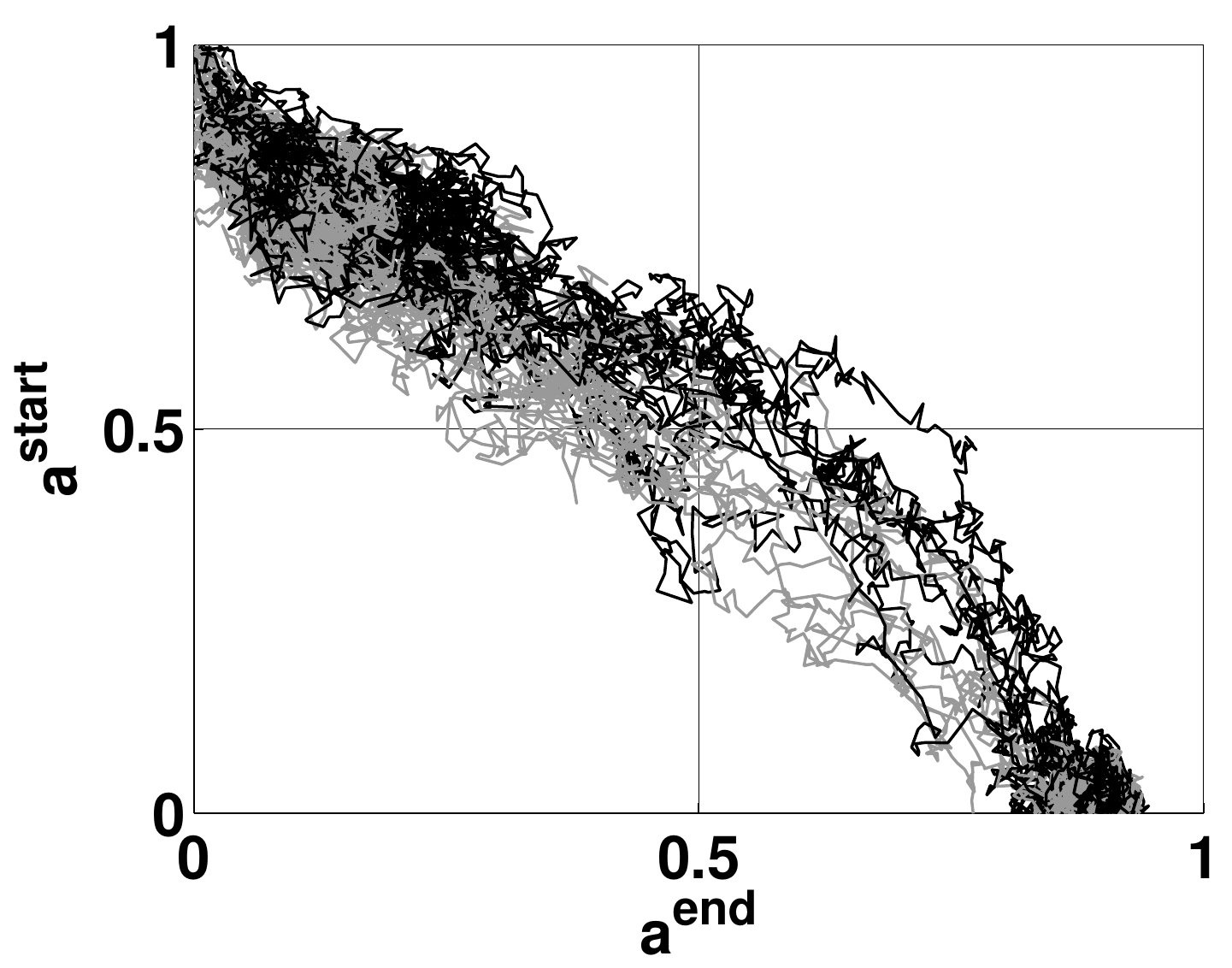}
\caption{\textbf{SI Figure 8. Fig 2C simulations.} Figure 2C simulation inset enlarged and with more trajectories.}\end{figure}

\begin{figure}
\includegraphics[width=15.8cm]{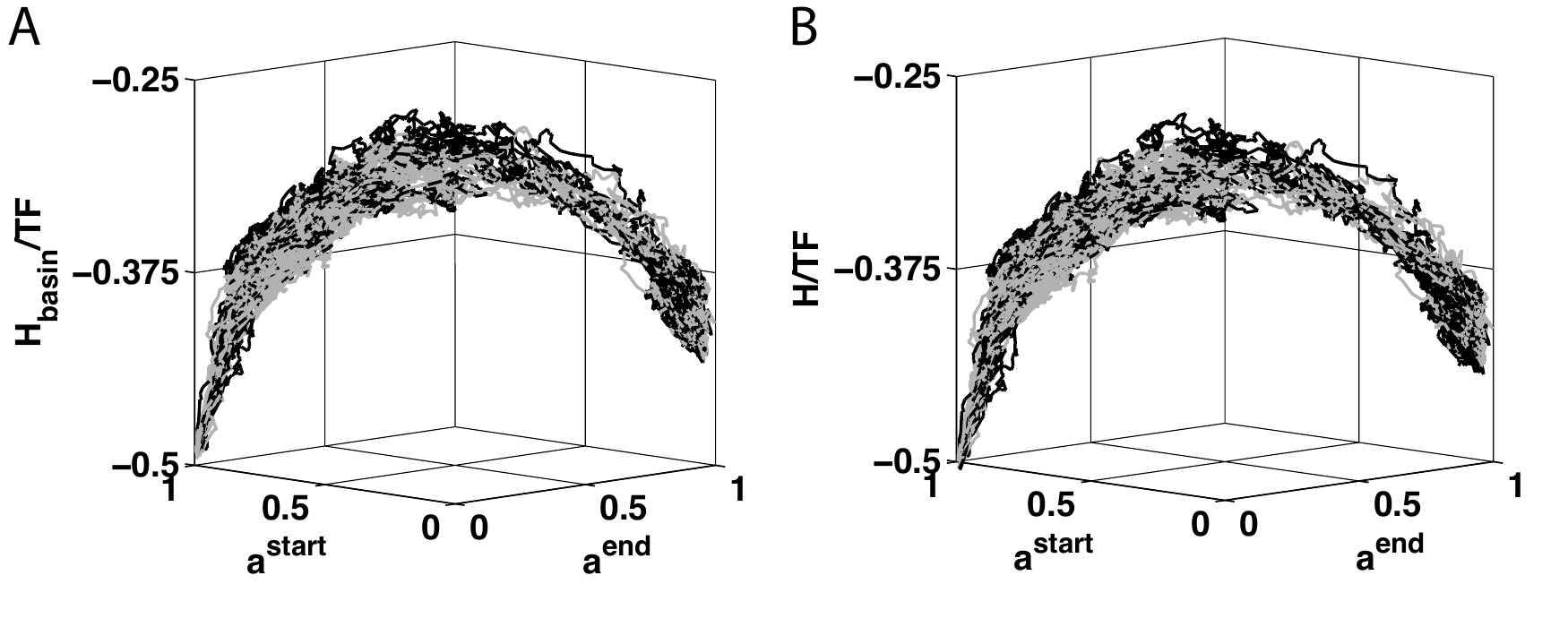}
\caption{\textbf{SI Figure 9. Fig 2D simulations.} A. Figure 2D simulation inset enlarged and with more trajectories. B. Figure 2D simulations including the small correction due to the culture term $H_{\text{culture}}$.}
\end{figure}

\begin{figure}
\includegraphics[width=10cm]{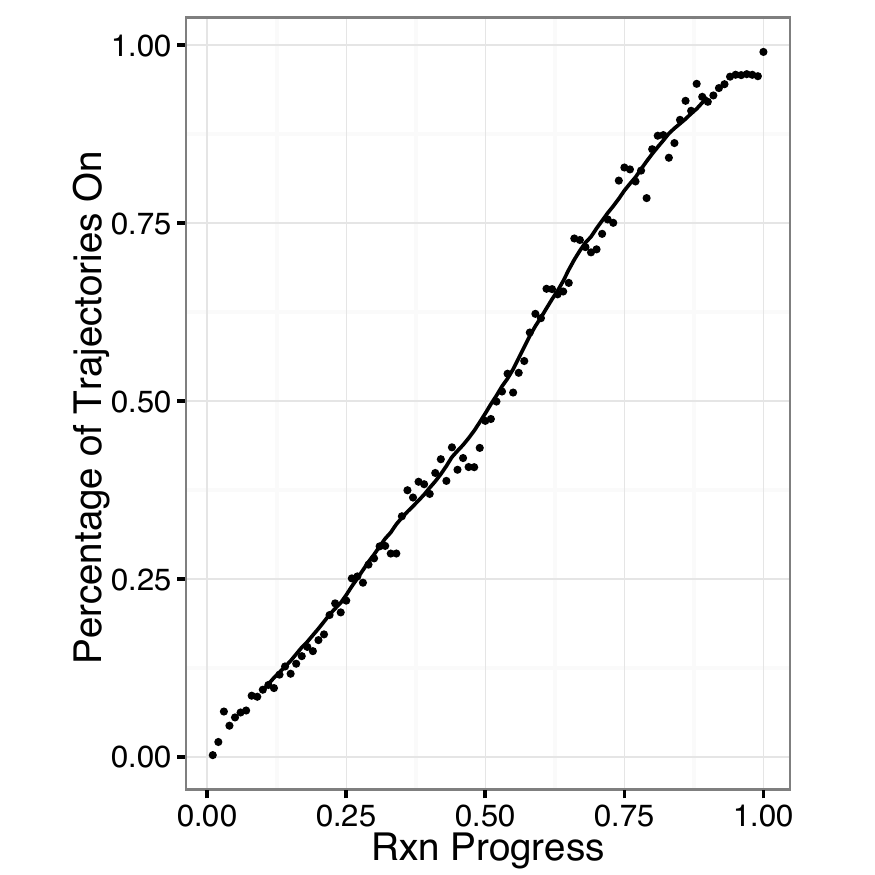}
\caption{\textbf{SI Figure 10.} Percentage of trajectories in which a gene is on vs reaction coordinate. Nanog is shown as an example where dots represent actual binned data, while the line is a 20 bin moving time average. This is just an example of the moving averages shown in Fig 3B.}\end{figure}

\begin{figure}
\includegraphics[width=\textwidth]{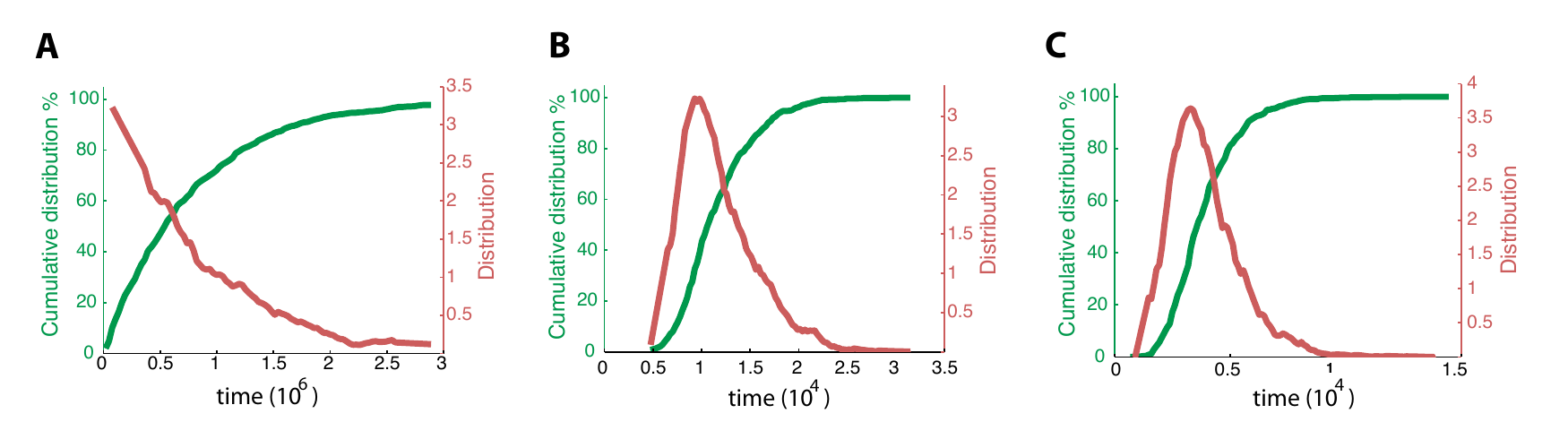}
\caption{\textbf{SI Figure 11.} A. Cumulative distributions of timing show that the early ($a^{end}=0$ to $a^{end}=0.3$), B. middle ($a^{end}=0.3$ to $a^{end}=0.7$), and C. late ($a^{end}=0.7$ to $a^{end}=0.8$) stages of reprogramming are respectively a Poisson, a narrowly peaked, and a narrowly peaked distribution.}\end{figure}

\end{document}